\documentclass[prb,multicol,twocolumn,showpacs,preprintnumbers,amsmath,amssymb,superscriptaddress]{revtex4}
\usepackage{graphicx}
\usepackage{dcolumn}
\usepackage{bm}

\begin{document}

\title{Coordinate shift in the semiclassical Boltzmann equation and the anomalous Hall effect} 
\author{N.A. Sinitsyn}
\affiliation{Department of Physics, University of Texas at Austin,
Austin TX 78712-1081, USA}
\author{Q. Niu} 
\affiliation{Department of Physics, University of Texas at Austin,
Austin TX 78712-1081, USA}
\author{A.H. MacDonald} 
\affiliation{Department of Physics, University of Texas at Austin,
Austin TX 78712-1081, USA}




\date{\today}

\begin{abstract}
Electrons in a crystal generically experience an anomalous coordinate shift (a side jump) when
they scatter off a defect.  
We propose a gauge invariant expression for the side jump associated with scattering between 
particular Bloch states.  Our expression for the side jump follows from the Born series expansion for the 
scattering T-matrix in powers of the strength of the scattering potential. 
Given our gauge invariant side jump expression, it is possible to construct a semiclassical
Boltzmann theory of the anomalous Hall effect which expresses all previously identified 
contributions in terms of gauge invariant quantities and does not refer 
explicitly to off-diagonal terms in the density-matrix response. 
\end{abstract}
\maketitle


\section { Introduction.}
 The near-equilibrium dynamics of a uniform system of
 classical charged particles in a weak electric field ${\bf E}$
 is described by the classical Boltzmann equation
 \begin{equation}
  \frac{\partial f_l}{\partial t} +e{\bf E}\cdot \frac{\partial f_{l} }{\partial {\bf k}}
  = -\sum_{l'}
   \omega_{ll'} (f_{l}-f_{l'})
 \label{beint}
 \end{equation}
 where $l=(\mu, {\bf k})$ is a combined index with ${\bf k}$ the momentum and  
 $\mu$ the label for some discrete internal degree-of-freedom,
 and 
$\omega_{ll'}$ is the scattering rate between unit momentum space volumes. 
 The momentum distribution function $f_{l}$ can be written as the sum of its equilibrium value $f_{0}(\epsilon_l)$, where
   $\epsilon_l$ is the energy dispersion, and
 a nonequilibrium correction $g_{l}$, {\em i.e.} $f_{l}=f_{0}(\epsilon_l)+g_{l}$. 
 Eq.\ref{beint} can also be applied successfully to quantum systems in many instances.  
 In the semi-classical theory of electronic transport in a crystal for example,
 $\mu$ becomes the Bloch band index, $f_{l}$ can be interpreted as a probability density in
 phase space that has been coarse grained by constructing wave packets,
 and $f_{0}(\epsilon_l)$ is the Fermi distribution function.
 
 In spite of the classical form of Eq.\ref{beint}
 the scattering rate $\omega_{ll'}$ often has to be calculated purely quantum mechanically and is given by the golden rule expression 
 in terms of the $T$-matrix, which can in turn be written as a Born series in powers of disorder strength.
 
 The semiclassical Eq.\ref{beint} is very powerful since it automatically takes care of the 
 summation of various infinite series of Feynman diagrams that appear in quantum linear response theory,
 and it keeps the physical meaning of all terms transparent. 
 However, since the only role of the electric
 field in Eq.\ref{beint} is 
 to accelerate wave packets constructed from states within a single band, and the only role of impurities is to 
 produce incoherent instantaneous scatterings, it appears clear that this 
 approach must be often insufficient. 
 The best known example where Eq.\ref{beint} apparently fails is in evaluating 
 the anomalous Hall effect (AHE) in spin-orbit coupled ferromagnetic metals, which 
 can be dominated by an inter-band coherence response.

 The rigorous quantum mechanical theory of the anomalous Hall effect in this case has been constructed
by Kohn and Luttinger \cite{{KohnLutt},{Luttinger}}, who considered the equation of motion of the density-matrix in momentum space. They
 found that elements of the density matrix that are off diagonal in the band index (interband coherence
contributions) are induced both by an external electric field and by disorder.  These couple to
off diagonal elements of the velocity operator, thus contributing to the Hall current 
  in addition to the skew scattering
contribution \cite{{Mott},{Smit56}}, that can be completely explained in the framework of  Eq.\ref{beint}.

 The renewed semiclassical theory, based on wave packet equations \cite{Niu},
 provided a simple explanation of those contributions.
 In this theory the Berry phase changes the velocity of wave packets and leads to the
 so called intrinsic contribution \cite{Luttinger54,Niu,Jungwirth,Yao,Fong,SinAHE,Experiment} to the AHE.
  Another ingredient in the semiclassical theory is to consider the charge transport not only
  {\em  between} collisions with
 impurities but also the transport {\em during} collisions, namely the so called side jump effect \cite{Berger}.

 The side jump is the coordinate shift acquired  by a particle during the scattering event. Recently, it was shown that for a smooth impurity 
 potential it can be found by integrating wave packet equations over the scattering time \cite{SinAHE}.
 After the gauge invariant expression for the side jump is found, one can calculate the related drift 
 velocity and the anomalous contribution to the distribution function, which in addition to the solution of the
 Eq.\ref{beint} are sufficient to calculate the Hall current \cite{SinAHE}.

  The advantage of the semiclassical approach is in its simplicity. It operates only with gauge invariant quantities, such as 
  the side jump, the scattering rate, anomalous and usual velocities and the distribution function. All of them have a clear physical meaning.
 In contrast, the approach by Kohn and Luttinger is rather obscure. The main reason is that the off-diagonal
elements of the density matrix or velocity operator are not gauge invariant.  Many of the individual 
contributions to the Hall effect in the Kohn-Luttinger approach are expressed in terms of gauge-dependent quantities
which cannot have separate physical meaning.  This may be one reason why many authors cite this article,
but try to invent their own way to calculate the Hall current \cite{{Dugaev},{Adams},{Nozieres},{Bruno-Dirac}}.
Another alternative approach can be
found in work by Adams and Blount \cite{Adams}, and Nozieres \cite{Nozieres} and proceeds by projecting
all operators to a single band of interest, or in the case of band-degeneracies 
in semiconductors, to a subsystem of two degenerate bands \cite{Nozieres}.
For electrons in the conduction band of common semiconductors, the disorder potential then acquires form 
$V({\bf r})\rightarrow V'= V({\bf r}) + \alpha {\bf \sigma} \cdot {\bf k} \times {\bf \nabla} V({\bf r})$
Projection to the subsystem also modifies the coordinate
operator ${\bf r} \rightarrow i\partial/\partial\bf k  +{\bf A}$, where
${\bf A}$ is the Berry connection of the band(s) (see \cite{{Bliokh-review}, {Blount-review}} for reviews). One then can define 
gauge-dependent  {\em side-jump} velocities $-i[i\partial/\partial{\bf k}, V']$ and
$-i[ {\bf A},V']$. Although such a technique can lead
to the correct answer it is framed in terms of non-commuting coordinates,
and gauge-dependent velocities.  The projected theory is in practice useful only for semiconductors
with sufficiently smooth disorder potential. 
Hence its applicability is strongly restricted.

Unfortunately, the gauge invariant approach to calculate the anomalous shift proposed in Ref.[~\onlinecite{SinAHE}] is also restricted only
to very smooth disorder potentials; it is not even 
obvious whether it is equivalent to the Born approximation in the weak potential limit, because the Born and the adiabatic
approximations often have very different domains of validity. 
Hence it would be valuable to find an alternative approach that 
applies to weak disorder potentials of arbitrary range. 

In the present work we propose a gauge invariant expression for the side jump for an arbitrary type of a scattering 
that can be treated in the Born approximation. 
This allows to evaluate all the major contributions to the anomalous Hall effect that 
have been identified in work by Kohn and Luttinger, but now using only classical concepts, without explicit reference 
to elements of the density matrix or Green functions that are off diagonal in band index or to non-commuting 
coordinates.

\section { Gauge invariant coordinate shift.}
To define the coordinate shift (side jump) in terms of well defined quantities that 
can be evaluated using scattering theory, we assume that 
a long time before the scattering event the center of mass of the wave packet moves freely according to the law 
${\bf r_{c}}(t)_{t \rightarrow -\infty}=\delta {\bf r_{-\infty}} + {\bf v_{k}}t$ 
where $ {\bf v_{k}}$ is the velocity of the free wave packet and $t$ is time.
Suppose the wave packet scatters from an impurity with the center at ${\bf r}_0=0$.
Then, if the momentum changes to ${\bf k'}$, a long time after the scattering event the coordinate of the outgoing state
center of mass should behave for $t \rightarrow +\infty$ as 
${\bf r_{c}}(t)_{t \rightarrow +\infty}=\delta {\bf r_{+\infty}} + {\bf v_{k'}}t$.  
We define the scattering induced coordinate shift as
\begin{equation}
\delta {\bf r}_{{\bf k',k}} = \delta {\bf r_{+\infty}}-\delta {\bf r_{-\infty}}
\label{shift1}
\end{equation}
The naive treatment  of the Boltzmann equation (\ref{beint}) based only on the scattering rate disregards this coordinate shift. 
Below we construct
the theory that enables us to calculate the coordinate shift in the lowest nonzero Born approximation
and add it to the contributions to the Hall conductance captured by calculations based on Eq.\ref{beint}.
For simplicity, we will consider first transport in a single band and later generalize results to the multiple band situation.
Let $\psi_{{\bf k}}({\bf r},t)=(1/  (2\pi)^{D/2}) e^{i{\bf kr}-i\epsilon({\bf k})t} |u_{{\bf k}} \rangle$ be the Bloch 
state with a momentum-dependent periodic spinor $|u_{{\bf k}} \rangle$.
The naive expression for the coordinate shift,
$ \delta {\bf r}_{{\bf k',k}}=\langle u_{{\bf k'}}|  i\frac{\partial}{\partial {\bf k'}} |u_{{\bf k'}} \rangle - \langle u_{{\bf k}}|
i\frac{\partial}{\partial {\bf k}} |u_{{\bf k}} \rangle$
is gauge dependent, {\em i.e.} it changes under an arbitrary momentum dependent phase change for the periodic spinors $|u_{{\bf k}} \rangle$,
and cannot be correct in general.  We derive the correct form for this expression in the weak potential scattering limit.
To find the correct expression, we prepare the wave packet that approaches the impurity. 
The wave function of the wave packet is a superposition of
eigenstates of the unperturbed Hamiltonian 
$\psi_{{\bf k}}({\bf r},t)$ with the real-valued Gaussian envelope 
factor $w({\bf k}-{\bf k_0})$, centered near the average momentum ${\bf k_0}$.
\begin{equation}
\Psi_{{\bf k_0}} ({\bf r},t) = \int d{\bf k} \; w({\bf k}-{\bf k_0}) \psi_{{\bf k}} ({\bf r},t)
\label{wp}
\end{equation}
We assume a vanishing width of the wave packet in momentum space in the usual way.  Hence, when multiplied by a smooth function of momentum
the envelope function $w({\bf k}-{\bf k_0})$ can be treated as a delta-function. However, when multiplied by a true delta-function, it
is considered smooth, reflecting the finite width of the wave packet. Correspondingly, in coordinate space, the wave packet should
be considered as large in comparison with a lattice constant, but small compared to other length scales.  We can 
evaluate its charge center as follows: 
\begin{equation}
{\bf r}_c({\bf k_0})=\int d {\bf r} \Psi_{{\bf k_0}} ({\bf r},t)^* {\bf r}  \Psi_{{\bf k_0}} ({\bf r},t)
\label{chcen}
\end{equation}
We substitute (\ref{wp}) into (\ref{chcen}), then
notice that ${\bf r}e^{i{\bf kr}}=-i(\frac{\partial}{\partial {\bf k}} e^{i{\bf kr}} )$ and integrate by parts.
 Using the orthogonality of plane waves,
 $\frac{\Omega_0}{(2\pi)^D} \sum_{\vec R}  e^{i({\bf k_1}-{\bf k_2}){\bf r}}=\delta ({\bf k_1}-{\bf k_2})$, and assuming that the 
 periodic functions are normalized, $\langle u_{{\bf k}}| u_{{\bf k}} \rangle =1$, and then the delta-function-like
  properties of envelope functions we finally find that, before scattering, the center of mass of such an unperturbed
 wave packet moves according to the law
 \begin{equation}
 {\bf r}_c ({\bf k_0},t)_{t\rightarrow -\infty}={\bf v}_{{\bf k_0}}t + \delta {\bf r_{-\infty}} = \frac{\partial \epsilon({\bf k_0})}{\partial {\bf k_0}} t + \langle u_{{\bf k_0}}|
 i\frac{\partial }{\partial {\bf k_0}} u_{{\bf k_0}} \rangle. 
 \label{chcen1}
 \end{equation}
($\Omega_0$ is the unit cell volume and $\sum_{\vec R}$ is a sum over lattice vectors.) 
Now  consider how the state which initially coincides with the Bloch state $\psi_{{\bf k}}({\bf r},t)$
moves under the  influence of a weak potential of an impurity $V({\bf r})$.
The solution of the Schr\"odinger equation can be written in terms of the eigenvectors of the unperturbed Hamiltonian $\psi_{{\bf k'}}({\bf r},t)$ as
\begin{equation}
\psi^{out}_{{\bf k}}({\bf r},t)=\int d {\bf k'} C({\bf k'},t) \psi_{{\bf k'}}({\bf r},t)
\label{split}
\end{equation}
To lowest order in the strength of the potential, perturbation theory leads to the following expression for time-dependent coefficients 
$ C({\bf k'},t)$ (see for example Eq. 19.9 in Ref.(~\onlinecite{Scattbook})): 
\begin{equation}
 C({\bf k'},t)=-iV_{{\bf k',k}} \int_{-\infty}^t e^{i(\epsilon({\bf k'}) -\epsilon ({\bf k}))t'} dt' + \delta ({\bf k'}-{\bf k})
\label{ckt}
\end{equation}
where $V_{{\bf k',k}} = \langle \psi_{{\bf k'}}({\bf r})|\hat{V}|\psi_{{\bf k}}({\bf r}) \rangle$ is the matrix element of the 
disorder potential between
two eigenstates of the unperturbed Hamiltonian.
Higher order terms can be incorporated into the above formula by substituting the 
$T$-matrix instead of the disorder potential matrix elements (see for example Eq. 19.10 in Ref.(~\onlinecite{Scattbook})).
\begin{equation}
 C({\bf k'},t)=-iT_{{\bf k',k}} \int_{-\infty}^t e^{i(\epsilon({\bf k'}) -\epsilon ({\bf k}))t'} dt' + \delta ({\bf k'}-{\bf k})
\label{ckt1}
\end{equation}
The time integral in (\ref{ckt}) is formally divergent, reflecting the fact that for infinite interaction time the initial state is completely destroyed.
We add a regularizing factor in the exponent 
$ e^{i(\epsilon({\bf k'}) -\epsilon ({\bf k}))t'} \rightarrow  e^{i(\epsilon({\bf k'}) -\epsilon ({\bf k}))t' -(\eta sign(t'))t'} $ to limit  
the effective finite time of interaction. Performing the integration in (\ref{ckt}) taking the limit $\eta \rightarrow 0$ after 
$t \rightarrow +\infty$ 
we thus find that at large positive time (see for example Eq. 19.60 in Ref.(~\onlinecite{Scattbook}))
\begin{equation}
 C({\bf k'},+\infty)=c({\bf k',k})+\delta ({\bf k'}-{\bf k})
\label{ckt2}
\end{equation}
where
\begin{equation}
c({\bf k',k})=-2\pi iT_{{\bf k',k}} \delta(\epsilon({\bf k'}) -\epsilon ({\bf k})) 
\label{ckt222}
\end{equation}
For ${\bf k'} \ne {\bf k}$ the square of the amplitude $|c({\bf k',k})|^2$ is the scattering probability from the state with momentum  ${\bf k}$ into the one with momentum
${\bf k'}$.  Due to the delta-function in (\ref{ckt222}), the expression $|c({\bf k',k})|^2$, should be understood in the regularized sense {\em i.e.} 
assuming that $\eta$ is small but finite.
Given these standard results from time-dependent perturbation theory, we can reconstruct the state of the wave packet after scattering
\begin{equation}
\Psi^{out}({\bf r},t) = \int d{\bf k} \; w({\bf k}-{\bf k_0}) \psi^{out}_{{\bf k}} ({\bf r},t)
\label{psip}
\end{equation}
where $\psi^{out}_{{\bf k}} ({\bf r},t)$ is given in (\ref{split}).
The average coordinate of the center of mass of the final state can be calculated by the same steps as used for the ingoing wave packet.
We find that
\begin{eqnarray}
{\bf r_c}(t)_{t\rightarrow +\infty} &=& \int d{\bf r} \left( \Psi^{out}({\bf r},+\infty)\right)^* {\bf r} \Psi^{out}({\bf r},+\infty) \nonumber \\
 &=& {\bf r_{+\infty}}^I + 
{\bf r_{+\infty}}^{II} + {\bf r_{+\infty}}^{III}
\label{finr}
\end{eqnarray}
where 
\begin{equation}
{\bf r_{+\infty}}^I = \int d {\bf k'} |C({\bf k'},+\infty)|^2 \left(   {\bf v_{k'}}t + 
 \langle u_{{\bf k'}}|
 i\frac{\partial }{\partial {\bf k'}} u_{{\bf k'}} \rangle  \right)
\label{r1}
\end{equation}
\begin{equation}
{\bf r_{+\infty}}^{II} = \int d {\bf k'} \left( c({\bf k',k_0}) \right)^*   i\frac{\partial }{\partial {\bf k'}} c({\bf k',k_0})
\label{r2}
\end{equation}
\begin{equation}
{\bf r_{+\infty}}^{III} = \lim _{{\bf k_1,k_2} \rightarrow {\bf k_0}}  \left( 
 i\frac{\partial }{\partial {\bf k_1}} c({\bf k_1,k_2}) -i\frac{\partial }{\partial {\bf k_2}}c^*({\bf k_2,k_1}) \right)  
\label{r3}
\end{equation}
${\bf r_{+\infty}}^{III}$ originates from the delta-function in Eq. \ref{ckt1}. 
In what follows, we will restrict our calculations to the lowest nonzero order in the potential $V_{{\bf k',k}}$. 

The perturbation expansion of the $T$-matrix is well known
\begin{equation}
T_{{\bf k',k}} = V_{{\bf k',k}} +\sum_{{\bf k''}} \frac{V_{{\bf k',k''}} V_{{\bf k'',k}}}{\epsilon({\bf k''}) - \epsilon({\bf k}) +i \eta} + \ldots
\label{TT}
\end{equation}
It will be useful to represent the disorder potential matrix elements 
in the form 
\begin{equation}
V_{{\bf k',k}} = |V_{{\bf k',k}}| e^{iArg(V_{{\bf k',k}})}.
\label{VV}
\end{equation}
Then, taking into account that 
$\int d {\bf k'} \frac{\partial }{\partial {\bf k'}}| c({\bf k',k_0})|^2 =0$, the Eq. \ref{r2} can be rewritten as
\begin{equation}
{\bf r_{+\infty}}^{II} = - \int d {\bf k'} | c({\bf k',k_0})|^2 \frac{\partial }{\partial {\bf k'}}Arg(V_{{\bf k',k_0}})  
\label{r22}
\end{equation}
Substituting (\ref{TT}) into (\ref{r3}) and noting that
$-2\pi i V_{{\bf k',k}} \delta (\epsilon({\bf k'}) - \epsilon({\bf k})) \approx c({\bf k',k}) $ we find to the second order in $ V_{{\bf k',k}}$ that
\begin{equation}
{\bf r_{+\infty}}^{III} =  - \int d {\bf k'} | c({\bf k',k_0})|^2 \frac{\partial }{\partial {\bf k_0}}Arg(V_{{\bf k',k_0}}) + 
f({\bf k_0}) {\bf v_{k_0}}
\label{r33}
\end{equation}
where $f({\bf k_0})$ is some function, whose exact expression will not be needed.
The last term in (\ref{r33}) does not break any symmetry and can be interpreted as renormalizing the normal velocity of the part of wave packet that
did not change the direction of motion after interacting with impurity. In what follows, we will ignore it as it has no influence on the Hall current
at the leading order of perturbation theory. 
Combining the remaining nontrivial terms from (\ref{r1}), (\ref{r22}) and (\ref{r33}) we find the
coordinate of the wave packet center of mass is 
\begin{equation}
\begin{array}{l}
{\bf  r_c}(t)_{t\rightarrow +\infty}=\int d {\bf k'} \; | C({\bf k'},+\infty)|^2 (  {\bf v_{k'}}t + \\
\\+ \langle u_{{\bf k'}}|
 i\frac{\partial }{\partial {\bf k'}} u_{{\bf k'}} \rangle  -{\bf \hat{D}_{k',k_0}} Arg(V_{{\bf k',k_0}}) )
\end{array}
\label{ansh}
\end{equation}
where ${\bf \hat{D}_{k',k_0}} = \frac{\partial }{\partial {\bf k'}} +  \frac{\partial }{\partial {\bf k_0}}$
The coefficient $ | C({\bf k'},+\infty)|^2$ can be interpreted as the scattering probability into state ${\bf k'}$ from the initial
state ${\bf k_0}$.
Thus Eq. \ref{ansh} has a semiclassical meaning such that the average final coordinate is the sum over probabilities of final states multiplied the
corresponding coordinate shifts. Combining this result with the expression for initial coordinate of the wave packet (\ref{chcen1}) one can read
the expression for the total anomalous shift corresponding to the 
scattering of the wave packet from the state with average momentum ${\bf k}$ into the one with ${\bf k'}$ in the lowest nonzero Born approximation.
\begin{equation}
 \delta {\bf r}_{{\bf k',k}} = \langle u_{{\bf k'}}| i\frac{\partial}{\partial {\bf k'}} u_{{\bf k'}} \rangle - 
 \langle u_{{\bf k}}| i\frac{\partial}{\partial {\bf k}} u_{{\bf k}} \rangle
 -  \hat{{\bf D}}_{{\bf k',k}} Arg(V_{{\bf k',k}} )
\label{dr1}
\end{equation}

Generalization to the multiple band case is simple. One should introduce the combined index $l=(\mu,{\bf k})$ in Eqs. \ref{split}, \ref{ckt}, \ref{TT}
and so on. Repeating the analogous steps we find that the expression for the anomalous coordinate shift after scattering from the state $l$ into the
state $l'$ is 
\begin{equation}
 \delta {\bf r}_{l'l} = \langle u_{l'}| i\frac{\partial}{\partial {\bf k'}} u_{l'} \rangle - 
 \langle u_{l}| i\frac{\partial}{\partial {\bf k}} u_{l} \rangle
 - \hat{{\bf D}}_{{\bf k',k}} Arg(V_{ l',l} )
\label{delr4}
\end{equation}

Equation  (\ref{delr4}) is the main result of this work. It provides a gauge invariant expression
 for the wave packet coordinate shift (side jump) which has
a clear semiclassical interpretation and is valid for an arbitrary impurity potential that can be treated in the Born approximation.

\section{Relation to Pancharatnam phase}
 Under the gauge transformation $| u_{l} \rangle
\rightarrow exp(i\phi_{\mu}({\bf k}))| u_{l} \rangle$  the argument of the potential operator matrix element in (\ref{delr4}) 
changes as $Arg(V_{ l',l} ) \rightarrow Arg(V_{ l',l} )+\phi_{\mu}({\bf k})-\phi_{\mu'}({\bf k'})$,
 which compensates noninvariance of other two terms.  
To make more incite in symmetries and gauge invariance of the side jump expression we consider the case when the periodic part of the  Bloch 
function is only momentum dependent. For simplicity we consider scatterings in the same band only, from radial spin-independent impurity potential.
Then matrix elements of the potential become $V_{{\bf k',k}} = V^{0} ({\bf k'}-{\bf k})  \langle u_{{\bf k'}}|u_{{\bf k}}\rangle $, where
$ V^{0} ({\bf k'}-{\bf k}) \sim \int d{\bf r} exp(-i({\bf k'}-{\bf k})\cdot {\bf r}) V({\bf r})$ and for radial
impurity $\hat{{\bf D}}_{{\bf k',k}} Arg(V^{0} ({\bf k'}-{\bf k} ))=0$. Thus (\ref{dr1}) simplifies
\begin{equation}
 \delta {\bf r}_{{\bf k',k}} = \langle u_{{\bf k'}}| i\frac{\partial}{\partial {\bf k'}} u_{{\bf k'}} \rangle - 
 \langle u_{{\bf k}}| i\frac{\partial}{\partial {\bf k}} u_{{\bf k}} \rangle
 -  \hat{{\bf D}}_{{\bf k',k}} Arg(\langle u_{{\bf k'}}|u_{{\bf k}}\rangle )
\label{delr44}
\end{equation}
Interestingly, in this case the side jump does not depend on the form of the scattering potential explicitly. 
In the case of  a very small scattering angle \cite{Aleiner} ($|{\bf k'}-{\bf k} |<< |{\bf k}|$) one can
 (for non-degenerate bands) disregard interband scattering 
  and make additional approximations valid up to the first order in 
  the small parameter $|{\bf k'}-{\bf k} |$, for example  
 $|u_{{\bf k'}}\rangle \approx |u_{{\bf k}}\rangle + ({\bf k'}-{\bf k})|\frac{\partial u_{{\bf k}}}{\partial {\bf k}}\rangle$
 and $|\langle u_{{\bf k'}}|u_{{\bf k}}\rangle| \approx 1$. 
 Substituting this in (\ref{delr44}) we find that up to 1st order in $|{\bf k'}-{\bf k} |$ the 
 anomalous shift is
 \begin{equation}
  \delta {\bf r}_{{\bf k' k}} \approx {\bf F} \times ({\bf k'}-{\bf k})
 \label{delr_smooth}
 \end{equation}
where $  F_k = \epsilon_{ijk} Im \left( \langle \frac{\partial u_{{ \bf k}}}{\partial  k_j}| \frac{\partial u_{{\bf k}}}{\partial k_i}\rangle\right)$.
By definition, ${\bf F}$ is the gauge invariant momentum space Berry curvature of the Bloch band. 
The result for the anomalous shift (\ref{delr_smooth}) 
coincides with the one derived in Ref.(~\onlinecite{SinAHE}) in the adiabatic approximation.  In the smooth 
potential limit the adiabatic and Born approximation results coincide.  

One can check that 
(\ref{delr44}) is related to the gauge invariant Pancharatnam phase $\Phi_{{\bf k'',k,k'}}$
\begin{equation}
 \delta {\bf r}_{{\bf k',k}} = -\left(\frac{\partial\Phi_{{\bf k'',k,k'}}}{\partial {\bf k''}} \right)_{{\bf k''} \rightarrow {\bf k}} -
\left(\frac{\partial \Phi_{{\bf k'',k,k'}}} {\partial {\bf k''}} \right)_{{\bf k''} \rightarrow {\bf k'}}
\label{delr444}
\end{equation}
where 
\begin{equation}
\Phi_{{\bf k'',k,k'}}
=Arg(\langle u_{{ \bf k''}} | u_{{\bf k}} \rangle \langle u_{{ \bf k}} | u_{{\bf k'}} \rangle \langle u_{{ \bf k'}} | u_{{\bf k''}} \rangle)
\label{berry1}
\end{equation}
\begin{figure}
\includegraphics[width=8cm]{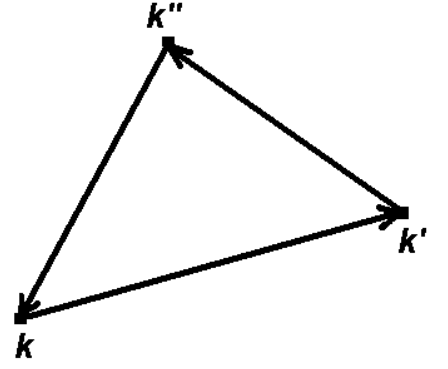}
\centering
\caption{Closed path in the momentum space representing the hopping amplitude of Eq. \ref{berry1}.}
\label{path}
\end{figure}
The Pancharatnam phase $\Phi_{{\bf k'',k,k'}}$ is the phase that would appear after the
hopping  over the closed path in the momentum space over the contour
 ${\bf k''}\rightarrow {\bf k} \rightarrow {\bf k'} \rightarrow {\bf k''}$ 
shown in Fig. \ref{path}.
One can demonstrate that the phase (\ref{berry1}) is also responsible for the skew scattering contribution. Taking the expression for $T$-matrix
(\ref{TT}) and calculating the scattering rate via the golden rule
$\omega_ {{\bf k',k}}=2\pi |T_{{\bf k',k}}|^2 \delta (\epsilon_{{\bf k}} -\epsilon_{{\bf k'}})$, we find for its asymmetric part the following
expression (see \cite{Luttinger} for the detailed derivation)
\begin{equation}
\omega^{(3a)}_ {{\bf k,k'}}=-(2\pi)^2  \sum_{{\bf k''}} \delta (\epsilon_{{\bf k}}
 -\epsilon_{{\bf k''}})  \delta (\epsilon_{{\bf k}} -\epsilon_{{\bf k'}}) 
 Im \left( V_{{\bf k,k'}} V_{{\bf k',k''}} V_{{\bf k'',k}} \right) 
\label{omasym1}
\end{equation}
here the superscript $(3a)$ means that this is the anti-symmetric part of the scattering rate calculated up to the order $V^3$ 
in the disorder potential.
 The nonzero value of the phase (\ref{berry1}) is crucial to make the product of three potential matrix elements in (\ref{omasym1}) nonzero. 
\begin{equation}
Im ( V_{{\bf k,k'}} V_{{\bf k',k''}} V_{{\bf k'',k}})\sim
Im( \langle u_{{ \bf k}} | u_{{\bf k'}} \rangle \langle u_{{ \bf k'}} | u_{{\bf k''}} \rangle \langle u_{{ \bf k''}} | u_{{\bf k}} \rangle)
\label{imvvv}
\end{equation}
\section{ Application to the AHE}

We now apply the side-jump expression (\ref{delr4}) to the AHE using ideas that have their roots in
earlier work that started from adiabatic approximations \cite{{Nozieres},{SinAHE}}.
The average side-jump velocity can be expressed in terms of the rate of transitions and 
the side-jump associated with a particular transition: 
\begin{equation}
\begin{array}{l}
{\bf v}_{l}^{sj} =\sum_{l'} \omega_{l'l} \delta {\bf r}_{l'l} =\\
\\
=\sum_{l'} 2\pi N \delta(\epsilon_l-\epsilon_{l'})
[ |V_{l'l}|^2 ( 
 \langle u_{l'}| i\frac{\partial}{\partial {\bf k'}} u_{l'} \rangle - \\
\\- \langle u_{l}| i\frac{\partial}{\partial {\bf k}} u_{l} \rangle  )
 -Im \left(   V_{ll'} \hat{{\bf D}}_{{\bf k' k}} V_{l'l} \right) ].
 \end{array}
\label{vsj}
\end{equation}
Here we have used lowest Born approximation expression $\omega_{l'l}=2\pi N \delta(\epsilon_l-\epsilon_{l'}) |V_{l'l}|^2 $, where 
$N$ is the impurity concentration.
A similar expression   can be found in the second part of Eq. 2.38 in
Ref.(~\onlinecite{Luttinger}). Luttinger called 
this velocity {\em the off-diagonal velocity} because its calculation involved interband matrix elements of the velocity operator. 
Our result (\ref{vsj}) is more general because we did not assume, as in Ref.(~\onlinecite{Luttinger}), that Bloch bands are nondegenerate and
that disorder potential is spin-independent. 

The side-jump velocity does not produce any contribution to the total current from the equilibrium distribution, but in an external
electric field a nonequilibrium correction to the distribution function appears.  This correction is well known \cite{{Luttinger}, {Loss}}
 $g_{l}  = \left( - \partial f_{0}/\partial \epsilon_l \right)eE_x|v_{l}| \tau_{\mu}^{||} cos (\phi)$, where
 $\tau_{\mu}^{||}$ is the transport life time defined as $1/\tau_{\mu}^{||}=\sum_{l'} \omega_{ll'}(1-(|v_{l'}|/|v_{l}|)cos(\phi - \phi'))$
 and  $\phi$ is the angle 
 between the velocity and the $\hat x$ direction, which we chose to be along the electric field.
 With this correction to the distribution, the side jump velocity leads to the current
 \begin{equation}
 {\bf J}^{sj} = e \sum_{l}  g_{l} {\bf v}_{l}^{sj}
 \label{jsj}
 \end{equation}
A second effect follows from the change of energy of the scattered particle under side-jump in the 
presence of an external electric field.
Since total energy is conserved, the scattered particle acquires additional kinetic energy
$\Delta \epsilon_{l'l}= \epsilon_{l'}-\epsilon_{l} = e {\bf E \cdot \delta r}_{l'l}$ in order to compensate the change in the 
potential energy in the electric field. 
The equilibrium distribution would then become unstable,
\begin{equation}
\begin{array}{l}
\frac{\partial f_l}{\partial t} = - \sum_{l'} \omega_{ll'}(f_{0}(\epsilon_{l})-f_{0}(\epsilon_{l'}))=\\
\\
=-\sum_{l'} \omega_{ll'}(-\frac{\partial f_{0}}{\partial \epsilon} \Delta \epsilon_{l'l}) \ne 0, 
\end{array}
\label{unst}
\end{equation}
unless compensated by an additional {\em anomalous} correction $g^a_{l}$ to the distribution function. Substituting in the 
collision term $f_{l}=f_{0}+g^a_l$ instead of $f_{0}$, in the stationary state we find the equation that determines $g^a_l$
\begin{equation}
\sum_{l'}\omega_{ll'}  \left( g^a_l - g^a_{l'} +\frac{-\partial f_0}{\partial \epsilon} e{\bf E} \cdot {\bf \delta r}_{l'l} \right) =0
\label{ganl}
\end{equation}
As in the case of the side jump velocity, we can find an analog of this equation in Luttinger's 
classic paper \cite{Luttinger}, however Luttinger
split the correction $g^a_l$ into nongauge invariant parts. One find that Eqs. 
3.21, 3.22, 3.23 and 3.16 of his work are equivalent to (\ref{ganl}).

In 2D the selfconsistent approach to calculate $g^a_l$ from (\ref{ganl}) is to look for a solution in the form  \cite{micr-cl}
$g^a_l(\phi)=\sum_{n} (g^a_{l})^{(n)} e^{in \phi} $.
  For the case of isotropic bands and isotropic scatterers
   one may calculate the Hall current at zero temperature without finding the expression for the full
  distribution function. Multiplying (\ref{ganl}) by $ev_{\mu} sin(\phi)$, 
   where $v_{\mu}$ is the Fermi velocity in the $\mu$-th band 
  and summing over ${\bf k}$, then
   taking into account that the Hall  current contribution from the band 
   $\mu$ is $I_{\mu}=e\sum_{{\bf k}} g_l^a v_{\mu} sin(\phi)$, we arrive at 
  the set of algebraic equations
  \begin{equation}
  I_{\mu}/ \tau_{\mu} -\sum_{\mu'} I_{\mu'} /\tau^c_{\mu,\mu'} + i_{\mu}=0
  \label{algebr}
  \end{equation}
  where $1/\tau_{\mu} \equiv \sum_{l'} \omega_{ll'}$, $1/\tau^c_{\mu,\mu'} \equiv \sum_{{\bf k}} \omega_{ll'} cos(\phi - \phi') v_{\mu}/v_{\mu'}$ and
  $i_{\mu} \equiv \sum_{{\bf k}} \frac{-\partial f_0}{\partial \epsilon_l} e^2 v_{\mu} sin(\phi) \sum_{l'}
   \omega_{ll'} {\bf E} \cdot \delta {\bf r}_{l'l}$.
 The final Hall current from the anomalous distribution is $J_y^{adist}=\sum_{\mu}I_{\mu}$.
  The above result can be simplified if direct transitions between different bands are for some reasons forbidden i.e.
   $\omega_{ll'}=\delta_{\mu \mu'} \omega^{\mu}_{{\bf k,k'}}$. Then, employing the symmetry of the problem one can derive a simple result
   \begin{equation}
   J_y^{adist}=J_y^{sj}
   \label{jj}
   \end{equation}
   where $J_y^{sj}$ is found in (\ref{jsj}). This equality has already been noticed in a less general context in Ref.(~\onlinecite{SinAHE}).

 In the semiclassical theory of the AHE the side jump effect is not the only disorder effect contributing to the
 Hall current. 
 In the weak disorder limit, the dominant contribution to the anomalous Hall effect 
 is rather due to skew scattering \cite{Mott,Smit56} which appears in the semiclassical Boltzmann equation 
 through\cite{Smit56,Luttinger,KohnLutt} the antisymmetric part of the Boltzmann equation 
 collision term kernel, $\omega_{l'l}-\omega_{ll'}$. 
 The first nonzero contribution to the asymmetric part of $\omega_{ll'}$ appears from the golden rule already in the  order $V^3$, however,
 that contribution is parametrically very different from others and one should prolong to calculate the asymmetric parts of
 $\omega_{ll'}$ up to $V^4$, including the localization correction, since
 corrections to Hall current due to $\omega_{ll'}$ in this order are parametrically similar with the side-jump contributions. 
 The technique of such 
 calculations is well known \cite{Luttinger}.
 The asymmetric scattering leads to the asymmetric correction to the distribution function $g_l^{\omega}$.
One can find it from a standard selfconsistent procedure, described in details, for example, in \cite{Loss} and the
corresponding current contribution is $J^{\omega}_y = e\sum_l g_l^{\omega}|v_l|sin(\phi)$. Thus skew scattering can be totally 
understood and calculated with the semiclassical Boltzmann equation. 
 
 Finally the important contribution to the AHE is the  
 Berry-phase contribution \cite{Luttinger54,Niu,Jungwirth,Yao,Fong,SinAHE,Experiment}.  
 This contribution is completely independent of scatterers 
 and is now often referred to as the intrinsic contribution to the anomalous Hall effect. 
 The intrinsic anomalous Hall effect
 has been evaluated explicitly in recent years for a variety of different ferromagnetic materials 
 using relativistic first principles 
 electronic-structure methods\cite{EbertReview,Lichtenstein,Anisotropic}.  
 It can be non-perturbative in character 
 because of band crossings,
 a property that partially explains the fact that it is often quantitatively important. 
 Although it is really an interband coherence effect, it can be captured in a semiclassical theory
 by working with modified Bloch bands \cite{Niu} that include band-mixing by the electric field 
 to leading order.  The end result 
 in this approach is the appearance of an anomalous velocity proportional to $|{\bf E}|$
 in addition to the usual velocity ${\bf v}_l=\partial \epsilon_l /\partial {\bf k}$.  The anomalous velocity
 ${\bf v}^{(a)}_l={\bf F}_l  \times e {\bf E}$ captures changes in the speed at which 
 a wave packet moves between  scattering events under the influence of the  external electric field
 only.  ${\bf F}_l$ is the Berry curvature of the band \cite{Niu}.
 The corresponding correction to the current is  ${\bf J}^{intrinsic}=e\sum_{l} f_{l}  {\bf v}^{(a)}_l$.

 Finally the total Hall current
in the transverse to the  electric field $y$-direction is
 \begin{equation}
 J^{total}_y=J^{intrinsic}_y+J^{sj}_y+J^{adist}_y+J^{\omega}_y
 \label{tot}
 \end{equation}

\section{Conclusions.}
In this work we demonstrated the importance  of the coordinate shift at a scattering event.
We found the general gauge invariant expression for this shift and related it to the
phases of the scattering 
$T$-matrix elements.
 We demonstrated that when equipped with this expression, the  semiclassical Boltzmann equation correctly reproduces
 all contributions to the AHE, that have been derived by Luttinger with a purely quantum mechanical approach. The existing
  alternative techniques inevitably have to deal either with adiabatic approximations or with 
  non-gauge invariant quantities like nondiagonal density matrix elements. 
Instead, the golden rule with our gauge invariant expression for the side jump and the semiclassical Boltzmann equation are sufficient
to derive the Hall current for arbitrary type of disorder.
%
Such calculations, though tedious, usually can be well 
automated with scientific software packages.  

Our conclusions about the role of the coordinate shift in the semiclassical Boltzmann equation are rather general and might be
important beyond the physics of the anomalous Hall effect.
Recently, Coulomb interactions and interactions with phonons and magnetic fields beyond conventional  approximations have been discussed
in the context of the Boltzmann equation \cite{{Coulumb_Bolt},{phonon-Bolt},{Glazov}}.
It would be interesting to trace the role of the coordinate shift in similar interacting systems.

\noindent
 {\it Acknowledgments}. The authors are grateful for useful discussions with 
Kentaro Nomura and Jairo Sinova. This work was supported by Welch Foundation, by DOE 
grant DE-FG03-02ER45958 and by NSF under the grant DMR-0115947.

\end{document}